\documentclass[12pt,a4paper]{article}

\usepackage{amsmath,amsfonts,amssymb,bm}

\usepackage{graphicx} 

\usepackage[hang,normalsize,bf]{subfigure} 

\begin{document}

\begin{center}

{\Huge \bf
Unintegrated CCFM \\ parton distributions \\
and transverse momentum  \\ of gauge bosons 
}

\vspace {0.6cm}

{\large \fbox{J. Kwieci\'nski} $^{1}$ and A. Szczurek $^{1,2}$}

\vspace {0.2cm}

$^{1}$ {\em Institute of Nuclear Physics\\
PL-31-342 Cracow, Poland\\}
$^{2}$ {\em University of Rzesz\'ow\\
PL-35-959 Rzesz\'ow, Poland\\}

\end{center}

\begin{abstract}
Transverse momentum distribution of gauge bosons $W^{\pm}$ and $Z^0$ is
calculated using unintegrated parton distributions
obtained by solving leading order CCFM equation(s) in the impact
parameter space.
For illustration we compare the results of the fixed-order collinear,
soft-gluon resummation and CCFM approaches.
A parameter of a nonperturbative form factor is adjusted to
the D0 data. In contrast to the collinear approach, the leading order
calculation based on the unintegrated CCFM parton distributions
generates transverse momentum distributions of gauge
bosons, which are almost consistent with experimental data provided
the parameter of the nonperturbative form factor is appropriately
adjusted.
The $W^+$-$W^-$ asymmetry is calculated as a function
of the boson rapidity and transverse momentum. Predictions are given
for RHIC and LHC.
\end{abstract}

\section{Introduction}

The production of gauge bosons at hadron colliders provides
fundamental tests of perturbative QCD. The understanding of transverse
momentum distribution is of particular interest in this context.
The data from the Tevatron collider cover regions of both small
and large gauge-boson transverse momenta ($p_t$)
\cite{D0_Z_2000,D0_W_2001,D0_WZ_2001}.
At large $p_t$ the standard fixed-order pQCD approach is adequate.
At small $p_t$ of vector bosons ($p_t \ll M_V$) one encounters
large logarithms of $M_V^2/p_t^2$. The leading term is of the form
\begin{equation}
\frac{d \sigma}{d p_t^2} \sim \frac{\alpha_s}{p_t^2} 
\ln \frac{M_V^2}{p_t^2} \; .
\label{large_log}
\end{equation}

The logarithms can be resummed to give a Sudakov form factor.
In recent years the resummation approach became the standard in
calculating transverse momentum distribution of gauge-bosons
\cite{CSS85,AK90,LY94,ERV97,EV98,KSV02}.
The most often used is the formalism proposed by Collins, Soper and Sterman
\cite{CSS85}. In the following we shall refer to it as CSS
or ``the b-space resummation approach''.
The b-space resummation approach is subjected to an ambiguity
in setting the matching condition between the resummed and fixed-order
results. Different prescriptions have been used throughout the literature.

The collinear parton approach was used by a few specialized group
to calculated many high-energy observables. 
Not all of them can be fully explained in the fixed-order collinear
parton model.
The $k_t$-factorization approach, using unintegrated parton distributions,
is an alternative approach.
In recent years there were many calculations of specific observables
using different sets of unintegrated parton distributions.
Mainly observables sensitive to unintegrated gluon distributions
have been discussed, like the dijet or heavy-quark pair production
in high-energy electron-proton scattering at HERA or the production of
heavy quark pairs in hadron-hadron collisions at high energy.
The existing unintegrated
gluon distribution differ one from another. In some approaches
(models) the unintegrated gluon distributions are
limited to the small-$x$ region by construction.  The unintegrated
(anti)quark distributions are very sparse and often not available.
Clearly, the program of a systematic study of the world data
in terms of unintegrated parton distributions (uPDFs) is at its beginning.\\

Gluon radiation along the ladder generates transverse momentum
of partons through the recoil effect.
Evolution of the unintegrated distributions is described then by the
so-called CCFM equation \cite{CCFM}.
The CCFM equation has a nice virtue that it interpolates
between the BFKL evolution at very small values of $x$ and the DGLAP evolution
at intermediate and large values of $x$. For many practical
applications it is very useful to solve the CCFM equation by means
of a Monte Carlo method \cite{JS01}.
The CCFM approach turns out very succesful in resolving some
high-energy puzzles \cite{small-x_collaboration,BJJPZ02,Jung1,Jung2}. 
In the so-called b-representation the CCFM equation for
b-dependent unintegrated distributions takes a simple diagonal
form \cite{CCFM_b1}. This approach can be generalized to include
in addition to unintegrated gluon distributions also unintegrated
(anti)quark distributions \cite{GKB03}.

The similarities and differences betwen the standard soft gluon
resummation and the CCFM resummation have been discussed recently
on the example of the $g g \rightarrow H$ fusion mechanism
\cite{GK03}.
It has been shown how the soft gluon resummation formulae can be
obtained as the result of an approximate treatment of the solution of
the CCFM equation in the so-called b-representation.

We shall show that in contrast to the collinear approach
the use of uPDFs leads to a non-$\delta$-like transverse momentum distribution
of the produced gauge bosons already in the leading order for matrix elements.
It is our aim here to calculate the transverse momentum distribution
of the charged gauge bosons based on the unintegrated parton
distribution originating from the numerical solution of the CCFM
equation.
This will be a first phenomenological test of the CCFM parton
distributions from \cite{GKB03}.
In addition, we wish to concentrate on the $W^+ - W^-$ asymmetry,
especially for proton-proton collisions at RHIC and LHC.
We shall show that this asymmetry strongly depends on the $\bar d -
\bar u$ asymmetry of parton distributions, providing a new interesting
tool to limit the latter somewhat better.

\section{Formalism}

In the fixed-order perturbative treatment the situation is as follows.
In the 0-th order collinear calculation the produced gauge bosons
carry no transverse momentum, i.e.
\begin{eqnarray}
&&\frac{d \sigma}{dy d^2 p_t} = \sigma_0^W
\sum_{qq'} |V_{qq'}|^2 \nonumber \\
&&\left[
  x_1 q_1(x_1,\mu^2) x_2 \bar q'_2(x_2,\mu^2)
+ x_1 \bar q'_1(x_1,\mu^2) x_2 q_2(x_2,\mu^2)
\right] \cdot \delta^2(\vec{p}_t) \; .
\label{standard_LO}     
\end{eqnarray}
For brevity, above we have introduced
$\sigma_0^W = \frac{2 \pi G_F}{3 \sqrt{2}}$.
The quark/antiquark momentum distributions are
evaluated at momentum fractions
$x_{1,2} = \frac{M_W}{\sqrt{s}} \exp(\pm y)$.

A nonzero transverse momentum of gauge
bosons is obtained only in the next-to-leading order calculation.
Then the gauge bosons with nonzero transverse momenta are produced
in the weak-QCD annihilation or  Compton processes
in association with a gluon (annihilation) or quark/antiquark
(Compton). The cross section associated with the 2 $\to$ 2 subprocess 
(1+2 $\to$ W+4) is given by the standard parton formula
\begin{equation}
\frac{d \sigma}{dy_3 dy_4 d^2 p_t} =
\frac{1}{16 \pi^2 {\hat s}^2}
\sum_{ij} x_1 p_1(x_1,\mu^2) x_2 p_2(x_2,\mu^2) \; \overline{\sum}
 |M_{ij}|^2 \;,
\label{standard_NLO}
\end{equation}
where the matrix elements squared, averaged over spins and
colors of incoming partons and summed over spins and colors of final
particles read:
\begin{equation}
\sum_{qq'}|M_{q \bar q' \to W g}|^2 = \pi \alpha_s(\mu_r^2)
 \sqrt{2} G_F M_W^2
|V_{qq'}|^2 \; \frac{8}{9} \frac{\hat{t}^2+\hat{u}^2+2 M_W^2 \hat{s}}
{\hat{t} \hat{u}}
\label{annihilation}
\end{equation}
for the annihilation and
\begin{equation}
\sum_{qq'}|M_{g q \to W q'}|^2 = \pi \alpha_s(\mu_r^2)
 \sqrt{2} G_F M_W^2
|V_{qq'}|^2 \; \frac{1}{3} \frac{\hat{s}^2+\hat{u}^2+2 M_W^2 \hat{t}}
{-\hat{s}\hat{u}}
\label{annihilation}
\end{equation}
for the weak Compton subprocess.
In the equations above $V_{qq'}$ are elements of the
Cabibbo-Kobayashi-Maskawa matrix.
The parton distributions are evaluated at
$x_{1,2} = \frac{1}{\sqrt{s}} \cdot
\left( m_{t,W} \exp(\pm y_1) + p_t \exp(\pm y_2) \right)$ where
the transverse mass $m_{t,W} = \sqrt{M_W^2 + p_t^2}$.

The poles at $\hat{t}$ = 0 or $\hat{u}$ = 0 cause the theoretical
cross section to diverge as $p_t \rightarrow$ 0 and exceeds
the experimental data at $p_t <$ 5 GeV (see Fig.\ref{fig:1st_order}).
Virtual corrections to
$q \bar{q}' \to W$ contribute only at $p_t$ = 0
($d \sigma^W/d p_t^2 \propto \delta(p_t^2)$), ensuring that
the transverse momentum integrated cross section is finite, while leaving
problems at small finite transverse momenta.

In the CCS approach \cite{CSS85} the resummed cross section for
$W$ production can be written as
\begin{equation}
\begin{split}
\frac{d \sigma}{d y d^2 p_{t,W}} =
\sigma_0^W/(2 \pi)^2 \sum_{qq'} |V_{qq'}|^2
\int d^2 b \; J_0(p_t b) \; W^{NP}_{q\bar q'}(Q,b,x_1,x_2) \\
 x_1 \cdot \left[ q_1(x_1,\mu(b)) 
       + \frac{\alpha_s(\mu(b))}{2 \pi} C_{vc} q_1(x_1,\mu(b))
       + \frac{\alpha_s(\mu(b))}{2 \pi}(C_{q g} \otimes g_1)(x_1,\mu(b))
   \right]  \\
 x_2 \cdot \left[ \bar q'_2(x_2,\mu(b)) 
       + \frac{\alpha_s(\mu(b))}{2 \pi} C_{vc} \bar q'_2(x_2,\mu(b))
       + \frac{\alpha_s(\mu(b))}{2 \pi}(C_{\bar q' g} \otimes g_2)(x_2,\mu(b))
   \right]  \\
\exp\left[ \frac{1}{2} \left(S_q(b,Q) + S_{\bar q'}(b,Q) \right)
\right] \; ,
\label{CCS_approach}
\end{split}
\end{equation}
where the exponents in the Sudakov-like form factors read
\begin{equation}
S_q(b,Q) = S_{\bar q}(b,Q) =
- \int_{\bar \mu_{min}^2}^{Q^2} \frac{d \bar \mu^2}{\bar \mu^2}
\left[
\ln \left(\frac{Q^2}{\bar \mu^2} \right) A_q(\alpha_s(\bar \mu)) +
                                        B_q(\alpha_s(\bar \mu))
\right] \; .
\label{Sudakov}
\end{equation}
The coefficient $A$ and $B$ can be expanded in the series of
$\alpha_s$:
\begin{eqnarray}
A_q &=& 2 C_F + \frac{\alpha_s(\bar \mu)}{2 \pi} (...) + ... \; , \nonumber
\\
B_q &=& -3 C_F + \frac{\alpha_s(\bar \mu)}{2 \pi} (...) + ... \; .
\label{expansion}
\end{eqnarray}
The CSS formalism \cite{CSS85} leaves open the question of small $b$.
Different prescriptions have been proposed to treat this region
(see e.g.\cite{ERV97}).
The lower limit  of the integral in Eq.(\ref{Sudakov}) is usually taken
$\mu^2_{min} = \left( \frac{C_b}{b} \right)^2$, where $C_b$ = 2
$\exp(-\gamma_E ) \approx$ 1.1229. This prescription leads
to a kink for the Sudakov form factor if $C_b/b = Q$.
In order to allow a smooth dependence of the cross section on the gauge-boson
transverse momentum and to quarantee that the lower limit
is really lower than the upper limit, one could make the following
replacement
$\mu^2_{min} = \left( \frac{C_b}{b} \right)^2 \rightarrow 
\left( \frac{C_b}{b} \right)^2
\left[1 + C_b^2/(b^2 Q^2) \right]^{-1}$ (see e.g. \cite{ERV97}).
In order to quarantee that the scale of parton distribution does not take
unphysically small values we shall use in addition the following prescription:
\begin{equation}
\mu^2(b) = \mu^2_{min} + \mu_0^2 \; ,
\label{PDF_scale}
\end{equation}
where $\mu_0^2 \gg \Lambda_{QCD}^2$ is the starting value for
the QCD evolution.
For illustration, in the present paper we shall use easy to handle
the leading order parton distributions from Ref.\cite{GRV95}.

In the formalism of unintegrated parton distributions the nonzero
transverse momenta of gauge bosons are obtained already
in the leading order. The invariant cross section for inclusive
gauge boson production reads then as
\begin{eqnarray}
&\frac{d \sigma}{d y d^2 p_{t,W}} =
\sigma_0^W \sum_{qq'} |V_{qq'}|^2 \int
 \frac{d^2\kappa_1}{\pi} \frac{d^2\kappa_2}{\pi}
\;
\delta^2(\vec{p}_t-\vec{\kappa}_1-\vec{\kappa}_2)
\nonumber\\
&\left[ 
f_{q/1}(x_1,\kappa_1^2,\mu^2) f_{{\bar q'}/2}(x_2,\kappa_2^2,\mu^2) +
f_{{\bar q'}/1}(x_1,\kappa_1^2,\mu^2) f_{q/2}(x_2,\kappa_2^2,\mu^2)
\right] \; .
\label{uPDFs_LO}
\end{eqnarray}
In the equation above the delta function assures the conservation
of transverse momenta in the $q \bar q'$ fusion subprocess.
The momentum fractions are calculated as
$x_{1,2} = \frac{m_{t,W}}{\sqrt{s}} \exp(\pm y)$, where in contrast
to the collinear case $M_W$ is replaced by the transverse mass
$m_{t,W}$. In the case of the gauge boson production, the scale $\mu^2$
is taken as $M_W^2$ or $M_Z^2$.

We note that formally, as far as matrix elements are considered,
formula (\ref{uPDFs_LO}) is the
$k_t$-factorization counterpart of the collinear formula (\ref{standard_LO}).

Introducing unintegrated parton distributions in the space
conjugated to the transverse momenta \cite{CCFM_b1}
\footnote{In the present paper we shall use the notation $\tilde f$
instead of $\bar f$ as in Refs.\cite{GKB03,GK03} to avoid confusion
with antiquark distributions needed here.}

\begin{equation}
f_q(x,\kappa^2,\mu^2) = \frac{1}{2 \pi}
 \int  \exp \left( i \vec{\kappa} \vec{b} \right) \; 
\tilde f_q(x,b,\mu^2) \; d^2 b \; ,
\label{Fourier transform}
\end{equation}
and taking the following representation of the $\delta$ function
\begin{equation}
\delta^2(\vec{\kappa_1}+\vec{\kappa_2}-\vec{p}_t) =
\frac{1}{(2 \pi)^2} \int d^2 b \;
\exp \left[   
(\vec{\kappa_1}+\vec{\kappa_2}-\vec{p}_t) \vec{b}
\right] \; ,
\label{delta_representation}
\end{equation}
the formula (\ref{uPDFs_LO}) can be written in the equivalent way
\begin{eqnarray}
&\frac{d \sigma}{d y d^2 p_{t,W}} =
\sigma_0^W/\pi^2 \sum_{qq'} |V_{qq'}|^2
\int d^2 b \; J_0(p_t b) \nonumber \\
&\left[ 
\tilde{f}_{q/1}(x_1,b,\mu^2) \tilde{f}_{{\bar q'}/2}(x_2,b,\mu^2) +
\tilde{f}_{{\bar q'}/1}(x_1,b,\mu^2) \tilde{f}_{q/2}(x_2,b,\mu^2)
\right] \; .
\label{b-space_formula}
\end{eqnarray}
In the next section we show and discuss the results obtained with the
formula (\ref{b-space_formula}).
In the formulae for $Z^0$ boson production 
$|V_{q q'}|^2$ is replaced by
$\delta_{q q'} \frac{1}{2} (V_q^2 + A_q^2)$.

As already mentioned in the introduction, it is our intention here
to use uPDFs $\tilde{f}_q^{CCFM}(x,b,\mu^2)$ which fulfill b-space
CCFM equations \cite{CCFM_b1,CCFM_b2}.
However, the perturbative solutions $\tilde{f}_q^{CCFM}(x,b,\mu^2)$ do not
include nonperturbative effects such as, for instance,
intrinsic momentum distribution of partons in colliding hadrons.
In order to include such effects we propose to modify the perturbative
solution $\tilde{f}_q^{CCFM}(x,b,\mu^2)$
and write the modified parton distributions
$\tilde{f}_q(x,b,\mu^2)$ in the simple factorized form
\begin{equation}
\tilde{f}_q(x,b,\mu^2) = \tilde{f}_q^{CCFM}(x,b,\mu^2)
 \cdot F_q^{NP}(b) \; .
\label{modified_uPDFs}
\end{equation}
In the present study we shall use two different functional
forms for the form factor
\begin{equation}
F_q^{NP}(b) = F^{NP}(b) = \exp\left(- \frac{b^2}{4 b_0^2}\right) \;
\text{or} \; \exp \left( - \frac{b}{b_e} \right)
\label{formfactor}
\end{equation}
identical for all species of partons.
In Eq.(\ref{formfactor}) $b_0$ (or $b_e$) is the only free parameter.
In the next section we try to adjust this parameter to
the experimental data on transverse momentum distribution of $W^{\pm}$.

We wish to note an almost identical structure of the b-space resummation
formula (\ref{CCS_approach}) and our formula (\ref{b-space_formula}).
Assuming a factorizable form for the nonperturbative form factor
$W^{NP}_{q \bar q'}$:
\begin{equation}
W_{q \bar q'}^{NP}(Q,b,x_1,x_2) =
F_q^{NP}(Q,b,x_1) \cdot F_{\bar q'}^{NP}(Q,b,x_2) \; ,
\label{ff_factorization}
\end{equation}
the formulae (\ref{CCS_approach}) and (\ref{b-space_formula}) are
identical as far as the formal structure is considered,
if we define effective b-space unintegrated (anti)quark distributions
\begin{eqnarray}
{\tilde f}_{q_1}^{SGR}(x_1,b,Q^2) &=& \frac{1}{2}
 F_q^{NP}(Q,b,x_1) 
\left[ x_1 q_1(x_1,\mu(b)) + ... \right]
\exp \left( \frac{1}{2} S_q\left( b,Q \right) \right) \nonumber \\
{\tilde f}_{\bar q'_2}^{SGR}(x_2,b,Q^2) &=& \frac{1}{2}
 F_{\bar q'}^{NP}(Q,b,x_2)
\left[ x_2 \bar q'_1(x_2,\mu(b)) + ... \right]
\exp \left( \frac{1}{2} S_{\bar q'} \left( b,Q \right) \right)  \; .
\label{effective_uPDF}
\end{eqnarray}
The index $SGR$ above stands for ``soft-gluon resummation''.
The $k_t$-dependent unintegrated distributions of (anti)quarks corresponding
to the b-space resummation can be then obtained through the Fourier-Bessel
transform
\begin{equation}
f_q^{SGR}(x,\kappa^2,Q^2) = \int db b \; J_0(\kappa b)
\tilde{f}_q^{SGR}(x,b,Q^2).
\label{standard_uPDF_SGR}
\end{equation}

With the simple Ansatz (\ref{formfactor}) for $F_q^{NP}$
the whole $Q^2$ dependence resides exclusively in the Sudakov-like
form factor.

\section{Results}

\begin{figure}[htb] 
  \subfigure[]{\label{fig_d0_wpm_nlo_a}
    \includegraphics[width=7.0cm]{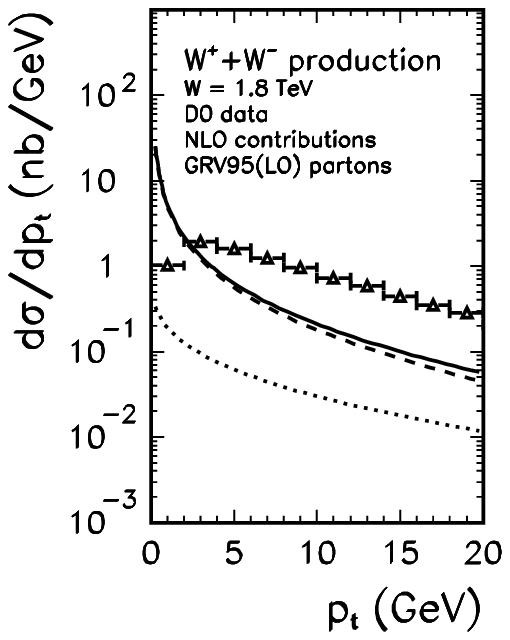}}
  \subfigure[]{\label{fig_d0_wpm_nlo_b}
    \includegraphics[width=7.0cm]{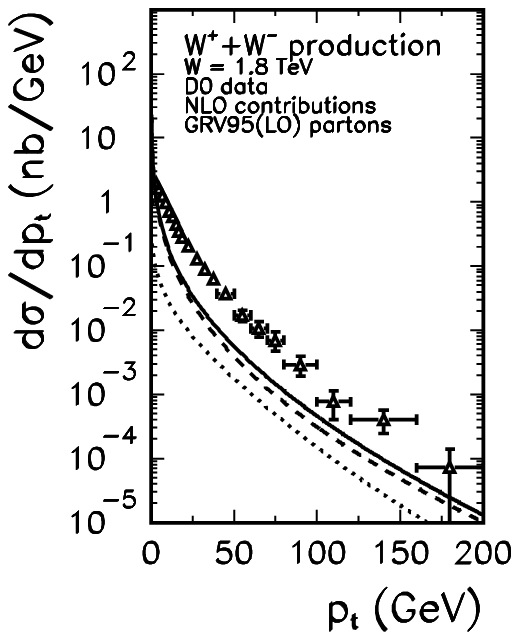}}
\caption{\it
Transverse momentum distribution of $W^+ + W^-$ at W = 1.8 TeV.
The curves represent the full first-order result (solid) and
the annihilation (dashed) and Compton (dotted) components separately.
The experimental data are taken from Ref.\cite{D0_WZ_2001}.
\label{fig:1st_order}
}
\end{figure}

For further reference in Fig.\ref{fig:1st_order} we present results
of the collinear NLO calculation. In this calculation the argument
of the running strong coupling constant $\mu_r^2 = M_W^2$ and the scale
argument is $\mu^2 = M_W^2$. In panel (a) we show the low-$p_t$ region,
whereas in panel (b) the whole measured region. The low-$p_t$ problems
are clearly visible. At the Fermilab energy the annihilation
contribution dominates over the Compton one in the entire
measured region. The 2 $\rightarrow$ 2 NLO contribution underestimates
the experimental data, especially in the interval 5 GeV $< p_t < $ 50 GeV.

\begin{figure}[htb] 
\begin{center}
\includegraphics[width=8cm]{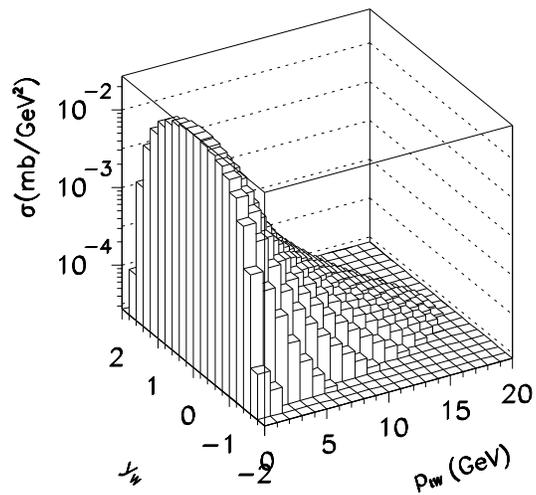}
\caption{\it
Two-dimensional map of the cross section for $W^+ + W^-$
production in proton-antiproton collisions at W = 1.8 TeV.
In this calculation $b_0$ = 1 GeV$^{-1}$ (Gaussian form factor).
\label{fig:sigma_ypt}
}
\end{center}
\end{figure}

In our CCFM calculation we have taken $\mu^2 = M_W^2$ as the
factorization scale in the b-space unintegrated parton distributions.
In Fig.\ref{fig:sigma_ypt} we present a two-dimensional map ($y,p_t$)
of the invariant cross section. Large transverse momenta of W bosons
are generated in the leading order calculations, which is due
to transverse momenta of fusing partons contained in unintegrated
(anti)quark distributions.

\begin{figure}[htb] 
  \subfigure[]{\label{fig_d0_wpm_ccfm_Gau}
    \includegraphics[width=7.0cm]{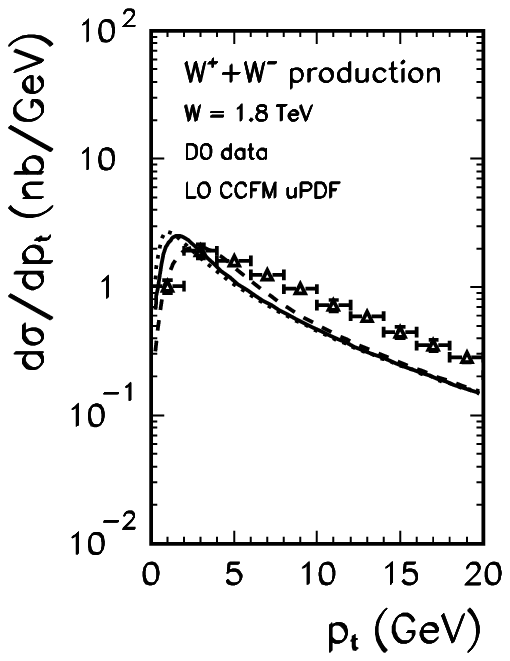}}
  \subfigure[]{\label{fig_d0_wpm_res_Gau}
    \includegraphics[width=7.0cm]{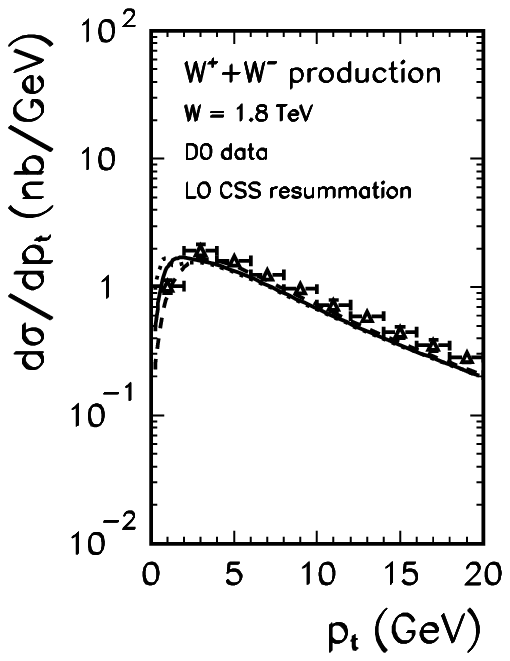}}
\caption{\it
Transverse momentum distribution of $W^+ + W^-$ at W = 1.8 TeV
within the formalism of the CCFM uPDF's (left panel) and soft-gluon
resummation (right panel) with Gaussian nonperturbative form factor.
The three curves correspond to different values of the form factor
parameter $b_0$ as explained in the text. The experimental data are
taken from Ref.\cite{D0_WZ_2001}.
\label{fig:D0_b0_W_gaussian}
}
\end{figure}

\begin{figure}[htb] 
  \subfigure[]{\label{fig_d0_wpm_ccfm_exp}
    \includegraphics[width=7.0cm]{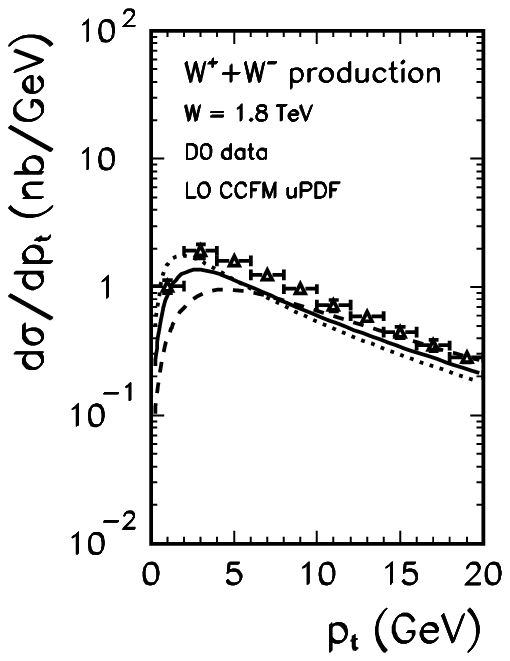}}
  \subfigure[]{\label{fig_d0_wpm_res_exp}
    \includegraphics[width=7.0cm]{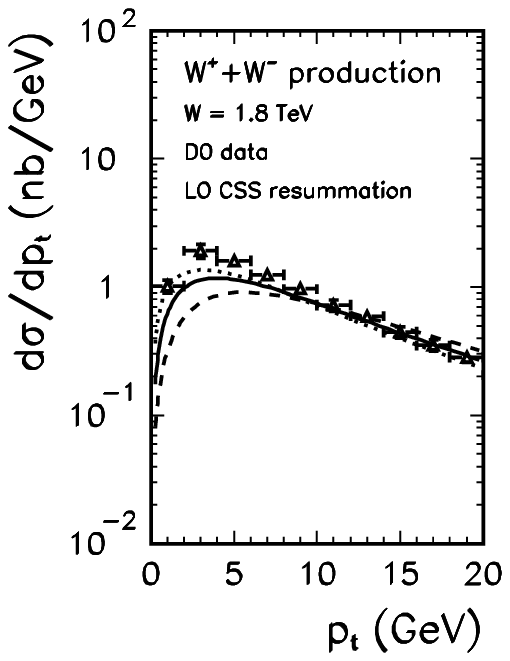}}
\caption{\it
Transverse momentum distribution of $W^+ + W^-$ at W = 1.8 TeV
within the formalism of the CCFM uPDF's (left panel) and soft-gluon
resummation (right panel) with exponential nonperturbative form factor.
The three curves correspond to different values of the form factor
parameter $b_e$ as explained in the text. The experimental data are
taken from Ref.\cite{D0_WZ_2001}.
\label{fig:D0_b0_W_exponential}
}
\end{figure}

In Fig.\ref{fig:D0_b0_W_gaussian} and
Fig.\ref{fig:D0_b0_W_exponential} we show transverse momentum distribution
(integrated over rapidities) of $W^{\pm}$ in proton-antiproton collissions
at Fermilab at W = 1.8 TeV for the CCFM parton distributions (left panel)
and standard soft-gluon resummation (right panel).
The three curves in Fig.\ref{fig:D0_b0_W_gaussian} show results
obtained according to formula
(\ref{b-space_formula}) with different values of the Gaussian form factor
parameter $b_0$: $b_0$ = 0.5 GeV$^{-1}$ (dashed), $b_0$ = 1.0 GeV$^{-1}$
(solid) and $b_0$ = 2.0 GeV$^{-1}$.
Similarly, the curves in Fig.\ref{fig:D0_b0_W_exponential}
correspond to $b_e$ = 0.5 GeV$^{-1}$ (dashed), $b_e$ = 1.0 GeV$^{-1}$
(solid) and $b_e$ = 2.0 GeV$^{-1}$ in the exponential form factor.
The results are overimposed on
the D0 collaboration data \cite{D0_WZ_2001} measured at Fermilab.
In order to convert the measured cross section ($e \nu$
channel) to the cross section for W production we have taken
the corresponding branching ratio $BR(e \nu)$ =0.1073
\cite{D0_WZ_2001}.
For the Gaussian form factor (see Fig.\ref{fig:D0_b0_W_gaussian})
somewhat better agreement with the data
is obtained with the CCS approach, while for the exponential
form factor (see Fig.\ref{fig:D0_b0_W_exponential}) it is the opposite.
One should remember, however, that the situation may change
if 2 $\rightarrow$ 2 NLO processes, neglected here, are included to
the CCFM formalism.
In the region of large momenta those contributions are expected to
be similar in size to the collinear NLO contributions shown in
Fig.\ref{fig:1st_order} (this is roughly the missing strength at $p_t$
= 20 GeV). In the CCS approach one somewhat arbitrarily
matches the soft-gluon resummation result with the fixed order collinear
result at the certain (chosen ad hoc) value of $p_t$. We hope that in
the formalism of unintegrated parton distributions in the
next-to-leading order approximation this can be done less arbitrarily.
The $D0$ data for W production are not precise enough to determine
the parameter $b_0$ (or $b_e$) of the nonperturbative form factor with high
precision. The low-$p_t$ $D0$ data for production of Z boson should
be better in this respect.
 
\begin{figure}[htb] 
\begin{center}
\includegraphics[width=8cm]{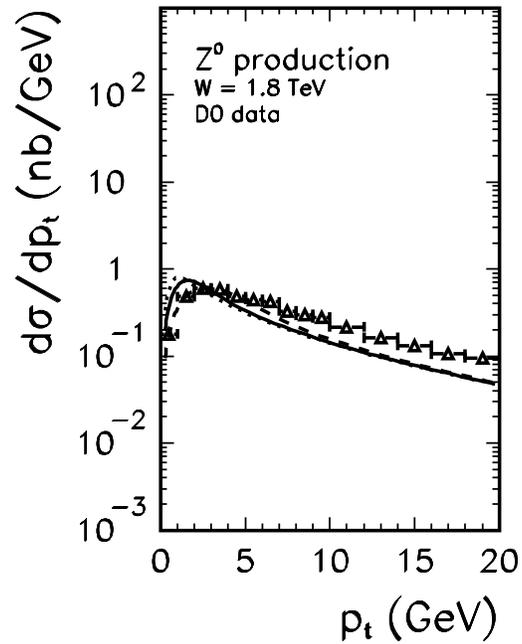}
\caption{\it
Transverse momentum distribution of $Z^0$ at W = 1.8 TeV.
The three curves correspond to different values of the Gaussian
form factor parameter $b_0$ as explained in the text.
The experimental data are taken from Ref.\cite{D0_Z_2000}.
\label{fig:D0_b0_Z}
}
\end{center}
\end{figure}

In Fig.\ref{fig:D0_b0_Z} we show transverse momentum distribution
(rapidity integrated) of $Z$ bosons in proton-antiproton collisions
at Fermilab at W = 1.8 TeV. As for the W production the theoretical
results are compared with the D0 experimental data \cite{D0_Z_2000}.
We have chosen the somewhat older data set \cite{D0_Z_2000}
instead of the most recent
one \cite{D0_WZ_2001} because of the binning with slightly better resolution
at small $p_t$. In order to convert the measured cross section ($e^+ e^-$
channel) to the cross section for Z production we have taken
the corresponding branching ratio $BR(e^+ e^-)$ =0.033632 \cite{D0_WZ_2001}.
The D0 data allow better extraction of the parameter $b_0$
of the nonperturbative form factor $F_{np}$. By comparison to the data
we find $b_0$ = 0.5--1.0 GeV$^{-1}$ (Gaussian)
as giving the best description of the low-$p_t$ $Z^0$-production data.

Comparing Fig.\ref{fig:1st_order} and Figs.\ref{fig:D0_b0_W_gaussian}
 and \ref{fig:D0_b0_W_exponential} one observes
that the leading-order CCFM approach generates more W's with
10 GeV $< p_t <$ 20 GeV than the standard next-to-leading order
collinear approach. The leading-order CCFM approach leaves some
room at $p_t >$ 5 GeV.
This is especially visible for the Gaussian form factor.
In this context it would be very interesting 
to calculate the cross sections with the help of CCFM uPDF's
up to the next-to-leading order.

\begin{figure}[htb] 
\begin{center}
\includegraphics[width=8cm]{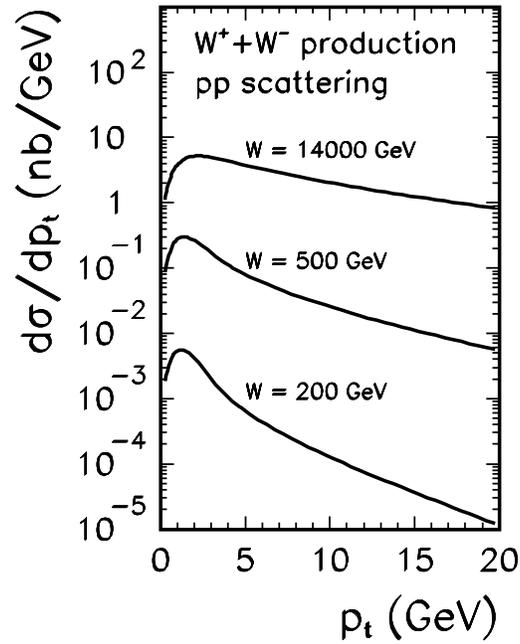}
\caption{\it
Transverse momentum distribution of $W^{\pm}$ in proton-proton
collisions for three different energies. In this calculation
-2 $< y <$ 2 and the parameter of the Gaussian form factor 
$b_0$ = 1 GeV$^{-1}$.
\label{fig:energy}
}
\end{center}
\end{figure}

In the future, the gauge bosons will be measured also in
proton-proton collisions at the BNL RHIC and at the CERN LHC.
As an example of our predictions in Fig.\ref{fig:energy} we present
the transverse momentum distribution of $W^{\pm}$ for three different
center-of-mass energies W = 200, 500, 14000 GeV.
In this calculation -2 $< y <$ 2 and the parameter $b_0$ was fixed
at 1 GeV$^{-1}$. The larger energy, the smaller-$x$ region of
(anti)quark distributions is probed. In principle, the form factor
in (\ref{formfactor}) can be not only a function of impact parameter $b$,
but also of partonic momentum fraction $x$. Having precise data at
different energies could help then to pin down the $x$ dependence
of the nonperturbative form factor. Other partonic processes
could also be helpful in this respect.

\begin{figure}[htb] 
  \subfigure[]{\label{fig_pp_y_0_rat}
    \includegraphics[width=7.0cm]{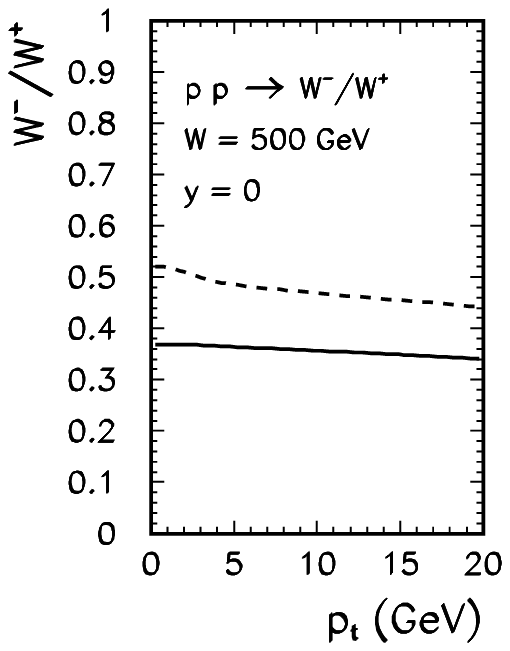}}
  \subfigure[]{\label{fig_pp_y_1_rat}
    \includegraphics[width=7.0cm]{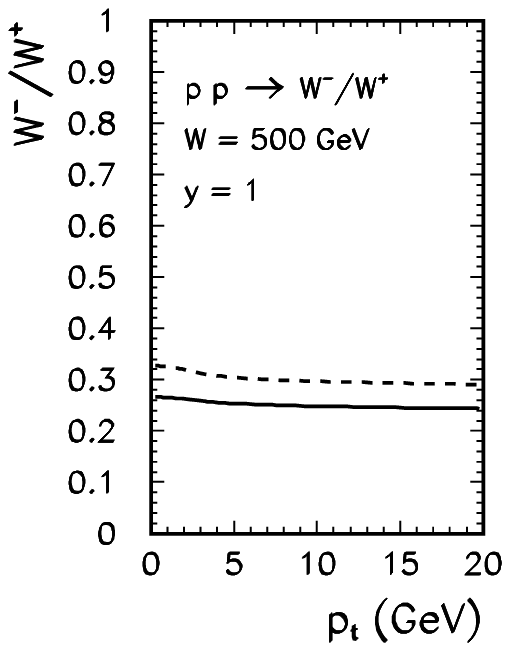}}
\caption{\it
The ratio $\sigma(W^-)/\sigma(W^+)$ for proton-proton
collisions at RHIC energy W=500 GeV
for y=0 (left panel) and y=$\pm$1 (right panel).
In this calculation the parameter $b_0$ of the nonperturbative
Gaussian form factor was fixed for $b_0$ = 1 GeV$^{-1}$.
The solid line corresponds to calculation with $\bar d - \bar u$
asymmetry while the dashed line to calculation with
symmetric sea.
\label{fig:rat_selected}
}
\end{figure}

\begin{figure}[htb] 
  \subfigure[]{\label{fig_asy_map_0}
    \includegraphics[width=7.0cm]{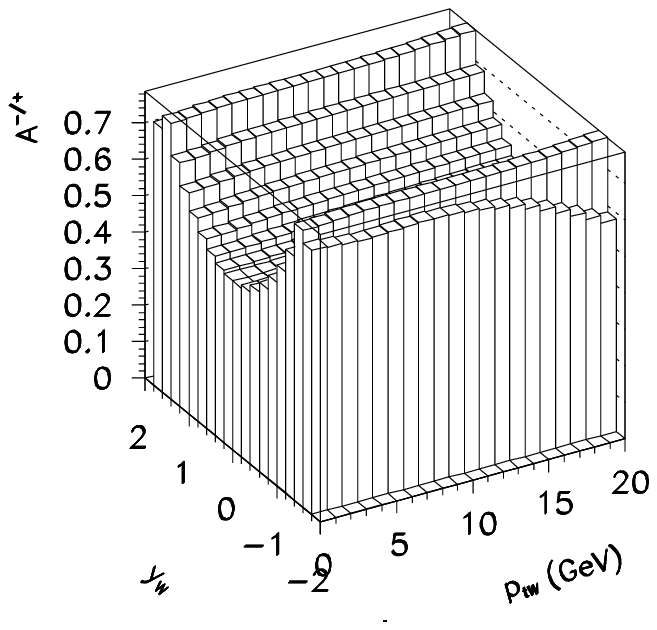}}
  \subfigure[]{\label{fig_asy_map_1}
    \includegraphics[width=7.0cm]{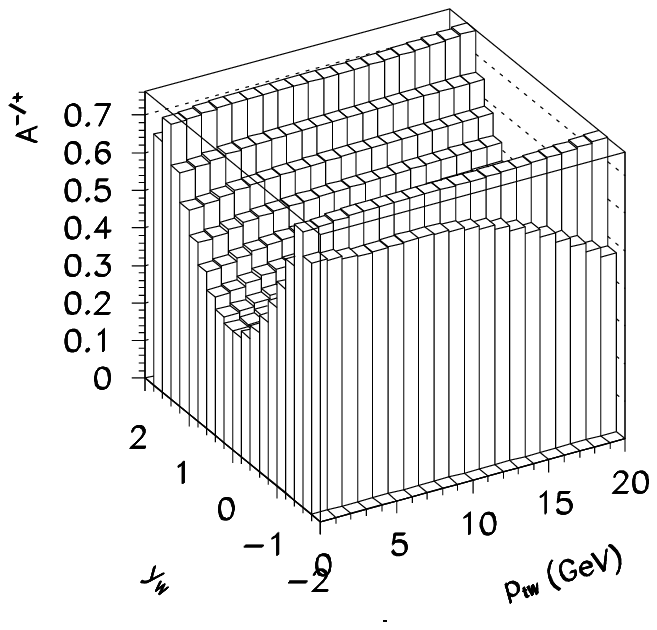}}
\caption{\it
Two-dimensional maps of the asymmetry $A_{\pm}$
for proton-proton
collisions at RHIC energies W=500 GeV for two cases:
in panel (a) with $\bar d - \bar u$ asymmetry and
in panel (b) without the asymmetry. In this calculation
the parameter $b_0$ of the nonperturbative Gaussian form factor
was fixed at $b_0$ = 1 GeV$^{-1}$.
\label{fig:asy_map}
}
\end{figure}

The precise determination of parton distributions is one of the
most important goals of high-energy physics.
It was suggested in Ref.\cite{SUHS97} that studies of the difference
in the production of $W^+$ and $W^-$ in proton-proton collisions
could put further constraints on the $\bar d - \bar u$ asymmetry as
obtained from other sources, like deep-inelastic inclusive and
exclusive scattering or Drell-Yan processes. In this illustrative
leading order calculations, however, one could not address the problem
of transverse momentum dependence of the $W^-/W^+$ ratio.
In contrast, it is very simple in our leading-order CCFM approach.
In Fig.\ref{fig:rat_selected} we present the ratio
as a function of W-boson transverse momentum at a typical RHIC
energy $W$ = 500 GeV for selected W-boson rapidities y=0 (left panel)
and y=$\pm$ 1 (right panel). The solid curve corresponds
to the calculation with the $\bar d - \bar u$ asymmetry, while
the dashed curve shows the result with
$\bar d -\bar u$ asymmetry switched off.
In general, the ratio $R=\sigma(W^-)/\sigma(W^+)$ is smaller if
the $\bar d-\bar u$ asymmetry is included. The relative effect is
larger for y=0. Fig. \ref{fig:rat_selected} illustrates
therefore the potential to study the $\bar d - \bar u$ asymmetry
in W-boson production at RHIC.
There is only a very weak dependence on W-boson transverse momentum.
In order to illustrate the dependence on W-boson rapidity
in Fig.\ref{fig:asy_map} we display also the two-dimensional maps ($y$,$p_t$)
of the asymmetry defined as
\begin{equation}
A_{\pm} = \frac{\sigma(W^+) - \sigma(W^-)}{\sigma(W^+) + \sigma(W^-)}
\label{asymmetry}
\end{equation}
for asymmetric (left panel) and symmetric (right panel)
antiquark distributions.
The asymmetry so-defined is an asymmetric function under the operation
$y \to -y$ for proton-antiproton collisions and a symmetric one for
proton-proton collisions.
The STAR collaboration intends to measure the W-boson production
in a few year perspective. It is not clear to us whether the map shown in
Fig.\ref{fig:asy_map} can be studied experimentally in the near future
at the Brookhaven RHIC.

\section{Conclusions}

The zeroth order collinear calculation leads to the $\delta$-like
distribution in the W-boson transverse momentum.
The 1-st order calculation overestimates the experimental data
at $p_t <$ 5 GeV and underestimates the data at $p_t >$ 5 GeV.
In contrast, already the leading order CCFM approach leads to a reasonable
transverse momentum distributions of W-bosons.
The small transverse momenta $p_t <$ 10 GeV are sensitive
to the nonperturbative physics embodied in our calculation
in the extra b-dependent form factors.
The parameter(s) of the form factor can be adjusted to
the experimental low-$p_t$ distributions provided the data are
sufficiently precise.
The simple leading order calculation leaves some room
for next-to-leading order contributions at $p_t >$ 5-10 GeV.

In the present study we have used very simple functional
forms of the nonperturbative form factors responsible for
intrinsic momentum distribution of partons in colliding hadrons.
It is not obvious a priori if these forms are sufficient to
describe the variety of data including
Drell-Yan dilepton production, prompt photon production,
heavy quark pair production, etc.
This will be a subject of the forthcoming research.

We have studied the dependence of the cross section ratio
$W^-/W^+$ in proton-proton collisions on transverse momentum
and rapidity of the W-boson.
We find a potential to study the $\bar d - \bar u$
asymmetry at BNL RHIC quantitatively.

There are many other possible ways to study the unintegrated
parton distributions further. For instance at RHIC energy
the topics to study can be jet or/and
particle production \cite{szczurek03}.

\vskip 1cm

{\bf Acknowledgments}
This paper was completed after Jan Kwieci\'nski passed away.
One of us (A.S.) is indebted to Krzysztof Golec-Biernat for a
discussion.



\end{document}